\documentclass[12pt]{article}
\usepackage{amssymb}
\usepackage{amsmath}

\makeatletter
\def\numberbysection{\@addtoreset{equation}{section}
 	\def\theequation{\thesection.\arabic{equation}}}
\makeatother

\numberbysection

\newcommand{\be}{\begin{eqnarray}}
\newcommand{\ee}{\end{eqnarray}}
\newcommand{\non}{\nonumber}
\newcommand{\tr}{\mathop{\rm tr}\nolimits}

\newcommand{\id}{\mathbb{I}}
\newcommand{\csch}{\mathop{\rm csch}\nolimits}
\newcommand{\sech}{\mathop{\rm sech}\nolimits}

\begin{document}

\begin{titlepage}
\strut\hfill UMTG--242
\vspace{.5in}
\begin{center}

\LARGE Completeness of the Bethe Ansatz solution\\
\LARGE of the open XXZ chain with nondiagonal boundary terms \\[1.0in]
\large Rafael I. Nepomechie \footnote{
    Physics Department, P.O. Box 248046, University of Miami,
    Coral Gables, FL 33124 USA}
and Francesco Ravanini \footnote{
    INFN Sezione di Bologna, Dipartimento di Fisica,
    Via Irnerio 46, 40126 Bologna, Italy}\\

\end{center}

\vspace{.5in}

\begin{abstract}
A Bethe Ansatz solution of the open spin-${1\over 2}$ XXZ quantum spin
chain with nondiagonal boundary terms has recently been proposed. 
Using a numerical procedure developed by McCoy {\it et al.}, we find
significant evidence that this solution can yield the complete set of
eigenvalues for generic values of the bulk and boundary parameters
satisfying one linear relation.  Moreover, our results suggest that
this solution is practical for investigating the ground state of this
model in the thermodynamic limit.
\end{abstract}
\end{titlepage}

\setcounter{footnote}{0}

\section{Introduction}\label{sec:intro}

The Bethe Ansatz solution of the open spin-${1\over 2}$ XXZ quantum
spin chain with diagonal boundary terms has long been known
\cite{ABBBQ, Sk}.  However, the case of nondiagonal boundary terms
\cite{dVGR} has resisted solution for many years (see, e.g.,
\cite{Bat}).  A Bethe Ansatz solution for the latter case has recently
been proposed in \cite{Ne1,Ne2} (see also \cite{CLSW}).  In terms of
the parameters introduced there, the Hamiltonian is given by
\be
{\cal H }&=& {1\over 2}\Big\{ \sum_{n=1}^{N-1}\left( 
\sigma_{n}^{x}\sigma_{n+1}^{x}+\sigma_{n}^{y}\sigma_{n+1}^{y}
+\cosh \eta\ \sigma_{n}^{z}\sigma_{n+1}^{z}\right)\non \\
&+&\sinh \eta \Big[ 
\coth \alpha_{-} \tanh \beta_{-}\sigma_{1}^{z}
+ \csch \alpha_{-} \sech \beta_{-}\big( 
\cosh \theta_{-}\sigma_{1}^{x} 
+ i\sinh \theta_{-}\sigma_{1}^{y} \big) \non \\
&-& \coth \alpha_{+} \tanh \beta_{+} \sigma_{N}^{z}
+ \csch \alpha_{+} \sech \beta_{+}\big( 
\cosh \theta_{+}\sigma_{N}^{x}
+ i\sinh \theta_{+}\sigma_{N}^{y} \big)
\Big] \Big\} \,,
\label{Hamiltonian}
\ee
where $\sigma^{x} \,, \sigma^{y} \,, \sigma^{z}$ are the usual Pauli
matrices, $\eta$ is the bulk anisotropy parameter, $\alpha_{\pm} \,,
\beta_{\pm} \,, \theta_{\pm}$ are boundary parameters, and $N$
is the number of spins. An unusual feature of this Bethe Ansatz
solution is that the boundary parameters must satisfy the linear relation
\be
\alpha_{-} + \beta_{-} + \alpha_{+} + \beta_{+} = \pm (\theta_{-} - 
\theta_{+}) + \eta k \,,
\label{constraint}
\ee
where $k$ is an even integer if $N$ is odd, and is an odd integer if
$N$ is even.  \footnote{An alternative solution was proposed in
\cite{XXZ} which does not require any constraint among the boundary
parameters.  However, that solution holds only for $\eta$ values
corresponding to roots of unity, and the Bethe Ansatz equations are
not of the conventional form.} The energy eigenvalues are given by
\cite{Ne2}
\be
E &=& \sinh^{2}\eta \sum_{j=1}^{M}{1\over \sinh u_{j}\ \sinh(u_{j}+\eta)}
+ {1\over 2} \sinh \eta\left( \coth \alpha_{-} + \tanh \beta_{-} +
\coth \alpha_{+} + \tanh \beta_{+} \right) \non \\
&+& {1\over 2} (N-1) \cosh \eta \,,
\label{energy}
\ee
where the Bethe roots $\{u_{j}\}$ satisfy the Bethe Ansatz equations
\be
{h(u_{j})\over h(-u_{j}-\eta)} = 
-{Q(u_{j}+\eta)\over Q(u_{j}-\eta)} \,, 
\qquad j = 1 \,, \ldots \,, M \,,
\label{openBAeqs}
\ee
where $h(u)$ is given by \footnote{This expression for $h(u)$ differs
from (3.26) in \cite{Ne2} by the factors $\kappa_{-} \kappa_{+}$ as
the result of working here with a rescaled transfer
matrix, as discussed further in 
Section \ref{subsec:transfer}.\label{footnote:rescale}}
\be 
h(u) &=& -\sinh^{2N}(u+\eta){\sinh(2u+2\eta)\over \sinh(2u+\eta)} 
\non \\
&\times& 4 \sinh(u + \alpha_{-}) \cosh(u + \beta_{-}) 
\sinh(u + \alpha_{+}) \cosh(u + \beta_{+}) 
\,, \label{hfinal} 
\ee
and $Q(u)$ is given by
\be
Q(u) = \prod_{j=1}^{M} \sinh(u - u_{j}) \sinh(u + u_{j} + \eta) \,,
\label{openQ}
\ee 
which satisfies $Q(u) = Q(-u-\eta)$. The Bethe Ansatz equations
assume a more symmetric form if expressed in terms of the shifted 
Bethe roots
\be
\tilde u_{j} \equiv u_{j} + {\eta \over 2} \,.
\label{shifted}
\ee 
Another unusual feature of this solution is that the number $M$ of
Bethe roots is fixed for given values of $N$ and $k$ (similar to the
case of the XYZ chain), and is given by
\be
M={1\over 2}(N-1+k) \,,
\label{Mvalue}
\ee
where $k$ is the integer appearing in (\ref{constraint}).

Several important issues were left unresolved in \cite{Ne2}.  In
particular, the Bethe Ansatz solution was obtained for the specific
values of the anisotropy parameter $\eta = i \pi/2\,, i \pi/4 \,,
\ldots$ (for which $q=e^{\eta}$ equals certain roots of unity).  The
solution was conjectured to hold for generic values of $\eta$, but
little direct evidence was given.  Also, because the value of the
integer $k$ in the constraint (\ref{constraint}) is tied to the number
of Bethe roots through (\ref{Mvalue}), it is evident that the
requirement of completeness should restrict the value of $k$. 
However, the problem of determining those restrictions remained
unsolved.

The purpose of this article is to address these and related questions. 
From numerical studies of chains with sizes up to $N=7$, we find
significant evidence that the Bethe Ansatz solution indeed holds for
generic values of $\eta$.  In particular, for generic values of both
bulk and boundary parameters, the Bethe Ansatz solution yields the
{\it complete} set of $2^{N}$ eigenvalues when $k=N+1$ (i.e., $M=N$). 
While it is gratifying to obtain all the eigenvalues, this result is
also disappointing, as it is impractical to satisfy the constraint
(\ref{constraint}) with $k=N+1$ in the thermodynamic ($N \rightarrow
\infty$) limit.  However, in practice one is interested primarily in
the lowest-lying levels; and we find significant evidence that the
Bethe Ansatz solution yields the {\it ground state} energy with just
$k=1$ (i.e., $M={1\over 2}N$) for $N$ even, and with $k=0$ (i.e.,
$M={1\over 2}(N-1)$) for $N$ odd.  We also investigate the special
case
\be
\alpha_{-}=-\alpha_{+} \,, \qquad 
\beta_{-}=-\beta_{+} \,, \qquad 
\theta_{+} = \theta_{-} = 0\,, \qquad N =  \mbox{odd} 
\label{specialcase}
\ee
considered in \cite{Ne1}, and find evidence that the Bethe Ansatz
solution gives the {\it complete} set of eigenvalues with $k=0$.

The outline of this article is as follows.  In Section
\ref{sec:generic}, we consider the Bethe Ansatz solution for generic
values of both bulk and boundary parameters.  We describe an ingenious
procedure, which was pioneered by Barry McCoy and his collaborators,
for addressing the problem of completeness; and we present the results
we have obtained using this procedure.  In Section \ref{sec:special}
we consider the special case (\ref{specialcase}).  We conclude with a
summary of our main results in Section \ref{sec:conclude}.

\section{Generic case}\label{sec:generic}

We begin by addressing the following question: given a value of $N$,
what value of $k$ is needed to obtain from the Bethe Ansatz solution
(\ref{constraint}) - (\ref{Mvalue}) the complete set of energy
eigenvalues?  To clarify the meaning of this question, let us make the
elementary observation that the number of Bethe roots must satisfy $M
\ge 0$; hence, (\ref{Mvalue}) implies that $k$ must satisfy $k \ge
1-N$.  The minimum value $k=1-N$ corresponds to zero Bethe roots, and
therefore, to only {\bf one} eigenvalue.  (See Eq.  (\ref{energy}).) 
Since a chain with $N$ spins has $2^{N}$ eigenvalues (which are
distinct for generic values of parameters), this minimum value of $k$
can give a complete set of eigenvalues only for $N=0$. 
\footnote{Although the Hamiltonian (\ref{Hamiltonian}) makes sense
only for $N \ge 2$, the transfer matrix (which is described below) is
well-defined even for $N=0$.} We wish to determine, for higher values
of $N$, the value(s) of $k$ needed to obtain from the Bethe Ansatz
solution the complete set of $2^{N}$ eigenvalues.

Such questions of completeness are notoriously difficult to address,
even numerically.  Indeed, since there is no known systematic way of
solving Bethe Ansatz equations, it is not possible to decide
unequivocally when one has found all the solutions of those equations. 
Fortunately, there does exist a systematic method, exploited by McCoy
and his collaborators (see, e.g. \cite{ADM, FM}), of determining the
Bethe roots corresponding to a given eigenvalue.  Since, for small
values of $N$, the eigenvalues can be computed by direct
diagonalization, this method can be used to determine whether the
Bethe Ansatz solution reproduces all the known eigenvalues.  (Since in
this approach one does not actually ``solve'' the Bethe Ansatz
equations, the possibility remains open that those equations may admit
additional solutions which do not correspond to actual eigenvalues. 
We shall return to this point later in Section \ref{subsec:massless}.)

This method, to which we refer as `McCoy's method', actually makes use
of the full transfer matrix of the model, rather than the Hamiltonian. 
Hence, we now briefly review its construction.

\subsection{Transfer matrix}\label{subsec:transfer}

The transfer matrix for the open chain (\ref{Hamiltonian}) is 
constructed according to Sklyanin's recipe \cite{Sk} from the 
$R$ matrix 
\be
R(u) = \left( \begin{array}{cccc}
	\sinh  (u + \eta) &0            &0           &0            \\
        0                 &\sinh  u     &\sinh \eta  &0            \\
	0                 &\sinh \eta   &\sinh  u    &0            \\
	0                 &0            &0           &\sinh  (u + \eta)
\end{array} \right) 
\label{bulkRmatrix}
\ee 
and the $2 \times 2$ nondiagonal matrices $K^{\mp}(u)$ whose components
are given by \cite{dVGR, GZ}
\be
K_{11}^{-}(u) &=& 2 \left( \sinh \alpha_{-} \cosh \beta_{-} \cosh u +
\cosh \alpha_{-} \sinh \beta_{-} \sinh u \right) \non \\
K_{22}^{-}(u) &=& 2 \left( \sinh \alpha_{-} \cosh \beta_{-} \cosh u -
\cosh \alpha_{-} \sinh \beta_{-} \sinh u \right) \non \\
K_{12}^{-}(u) &=& e^{\theta_{-}} \sinh  2u \,, \qquad 
K_{21}^{-}(u) = e^{-\theta_{-}} \sinh  2u \,,
\label{Kminuscomponents}
\ee
and
\be
K_{11}^{+}(u) &=& -2 \left( \sinh \alpha_{+} \cosh \beta_{+} \cosh (u+\eta) 
- \cosh \alpha_{+} \sinh \beta_{+} \sinh (u+\eta) \right) \non \\
K_{22}^{+}(u) &=& -2 \left( \sinh \alpha_{+} \cosh \beta_{+} \cosh (u+\eta) 
+ \cosh \alpha_{+} \sinh \beta_{+} \sinh (u+\eta) \right) \non \\
K_{12}^{+}(u) &=& -e^{\theta_{+}} \sinh  2(u+\eta) \,, \qquad 
K_{21}^{+}(u) = -e^{-\theta_{+}} \sinh  2(u+\eta) \,.
\label{Kpluscomponents}
\ee
The matrices $K^{\mp}(u)$ are equal to those appearing in \cite{Ne2}
divided by the factors $\kappa_{\mp}$, respectively.  This leads to
the rescaling of the transfer matrix already mentioned in Footnote
2.

The transfer matrix $t(u)$ is given by \cite{Sk}
\be
t(u) = \tr_{0} K^{+}_{0}(u)\  
T_{0}(u)\  K^{-}_{0}(u)\ \hat T_{0}(u)\,,
\label{transfer}
\ee
where the monodromy matrices are given by
\be
T_{0}(u) = R_{0N}(u) \cdots  R_{01}(u) \,,  \qquad 
\hat T_{0}(u) = R_{01}(u) \cdots  R_{0N}(u) \,,
\label{monodromy}
\ee
and $\tr_{0}$ denotes trace over the ``auxiliary space'' 0.
The transfer matrix has the important commutativity property
\be
\left[ t(u)\,, t(v) \right] = 0  \,,
\label{commutativity}
\ee 
and it ``contains'' the Hamiltonian (\ref{Hamiltonian}),
\be
{\cal H} = c_{1} {\partial \over \partial u} t(u) \Big\vert_{u=0} 
+ c_{2} \id \,,
\label{firstderivative}
\ee
where
\be
c_{1} &=& -{1\over 16 \sinh \alpha_{-} \cosh \beta_{-}
\sinh \alpha_{+} \cosh \beta_{+} \sinh^{2N-1} \eta 
\cosh \eta} \,, \non \\
c_{2} &=& - {\sinh^{2}\eta  + N \cosh^{2}\eta\over 2 \cosh \eta} 
\,,
\label{cees}
\ee 
and $\id$ is the identity matrix.
According to the Bethe Ansatz solution \cite{Ne2}, the eigenvalues
$\Lambda(u)$ of the transfer matrix are given by
\be
\Lambda(u) = h(u) {Q(u-\eta)\over Q(u)} 
+ h(-u-\eta) {Q(u+\eta)\over Q(u)}  \,,
\label{openeigenvalues} 
\ee
where $h(u)$ and $Q(u)$ are given by (\ref{hfinal}) and (\ref{openQ}),
respectively.  The formula (\ref{energy}) for the energy eigenvalues
follows directly from (\ref{firstderivative}) -
(\ref{openeigenvalues}).

\subsection{McCoy's method}\label{subsec:method}

To implement McCoy's method, it is more convenient to work with the
spectral parameter $x \equiv e^{u}$ and the anisotropy parameter $q
\equiv e^{\eta}$.  We denote by $t(x)$ the transfer matrix expressed
in terms of $x$, and similarly for other quantities.

McCoy's method consists of four steps: \footnote{We are grateful to B.
McCoy for explaining this procedure to us.}
\begin{enumerate}
    
    \item Fixing an arbitrary value $x_{0}$ of the spectral parameter,
   compute numerically the eigenvectors $|\Lambda \rangle$ of the 
   transfer matrix $t(x_{0})$. Due to the commutativity property of 
   the transfer matrix, the eigenvectors do not depend on the 
   spectral parameter. 
   
   \item Determine the eigenvalues $\Lambda(x)$ by acting with $t(x)$ on
   the eigenvectors $|\Lambda \rangle$.  Due to the commutativity
   property of the transfer matrix, these eigenvalues are Laurent
   polynomials in $x$.   
   
   \item Set 
   $Q(x) = \sum_{k=0}^{M} b_{k} \left( x^{2k} + (x q)^{-2k} \right)$
   (see (\ref{openQ})),
   and determine the coefficients $\{ b_{k} \}$ from the relation 
   (\ref{openeigenvalues}), i.e.,
   $\Lambda(x) Q(x) = h(x) Q({x\over q}) + h({1\over x q}) Q(x q)$.
   
   \item Factor the polynomials $Q(x)$, whose zeros $x_{j}$ are the 
   sought-after Bethe roots.

\end{enumerate}

Below we present the results that we have obtained using this method. 
We discuss separately the ``massless'' regime ($\eta$ is purely
imaginary) and the ``massive'' regime ($\eta$ is purely real).

\subsection{Massless regime}\label{subsec:massless}

For the case that the bulk anisotropy parameter $\eta$ is purely
imaginary, the transfer matrix is generally not Hermitian.  Hence, in
principle, it may have fewer than $2^{N}$ eigenvectors.  Nevertheless,
if the boundary parameters are suitably restricted, the transfer
matrix can be shown to be a normal matrix, i.e.,
\be
\left[ t(u)\,, t(u)^{\dagger} \right] = 0  \,,
\label{normal}
\ee 
which implies that it is unitarily diagonalizable.
Indeed, treating the spectral parameter $u$ as real, it is easy to see
that the $R$ matrix (\ref{bulkRmatrix}) satisfies
$R(u)^{\dagger}=-R(-u)$. Let us now restrict the boundary parameters
so that
\be
\alpha_{\mp} \,, \theta_{\mp} =  \mbox{purely imaginary}  \,, \qquad
\beta_{\mp} =  \mbox{purely real}  \,.
\label{masslessrestriction}
\ee 
The $K$ matrices (\ref{Kminuscomponents}),
(\ref{Kpluscomponents}) then obey similar relations $K^{\mp}(u)^{\dagger}
= -K^{\mp}(-u)$.  It follows that the transfer matrix obeys the simple
relation
\be
t(u)^{\dagger} = t(-u) \,.
\ee
Combining this result with the commutativity relation
(\ref{commutativity}) immediately yields the desired result
(\ref{normal}). Moreover, the conditions (\ref{masslessrestriction})
for the boundary parameters imply that  
the Hamiltonian (\ref{Hamiltonian}) is Hermitian.

In the numerical work which we present below, the values of the
bulk and boundary parameters are chosen as follows:
\be
\eta &=& 0.3 i \,, \quad \alpha_{+} = 0.75 i\,, \quad \beta_{+}= -0.5\,, 
\quad \theta_{+} = 0.8 i  \,, \non \\
\alpha_{-} &=& 0.25 i + (k-1)(0.3 i)\,, \quad \beta_{-}= 0.5 \,, 
\quad \theta_{-} = 0.1 i \,.
\label{masslessvalues}
\ee
This set of values satisfies the constraint (\ref{constraint}) for any
value of $k$, as well as (\ref{masslessrestriction}).

Table \ref{table:massless} shows, for values of $N$ ranging from 0 to
4, all the $2^{N}$ energy eigenvalues and the corresponding shifted
Bethe roots. 
Our main observation is that, for each value of $N$, the
corresponding value of $k$ is equal to $N+1$.  (We obtained similar
results for up to $N=7$, but we do not present the data here.)  For $k >
N+1$, the Bethe Ansatz also yields all $2^{N}$ energy eigenvalues. 
But for $1-N \le k < N+1$, the Bethe Ansatz does {\bf not} yield all
$2^{N}$ energy eigenvalues. \footnote{For the ``missing'' 
eigenvalues (i.e., those eigenvalues of the transfer matrix which are 
not given by the Bethe Ansatz solution), step 3 of McCoy's method 
fails: for such eigenvalues $\Lambda(x)$, there are no appropriate
polynomials $Q(x)$ which satisfy $\Lambda(x) Q(x) = h(x) Q({x\over q}) 
+ h({1\over x q}) Q(x q)$.} In other words, the minimum value of $k$
for which the Bethe Ansatz reproduces all the eigenvalues is $k=N+1$. 
We have observed this numerically for the choice of parameters
(\ref{masslessvalues}) with values of $N$ up to 7, and we conjecture
that it is true for generic values of the boundary parameters for all
$N$.

We further conjecture that for $k > N+1$, the Bethe Ansatz equations
(\ref{openBAeqs}) admit extraneous solutions, which do not correspond
to eigenvalues of the transfer matrix.  We have verified this
numerically for $N=0$ with $k=3$.  Indeed, with the choice of boundary
parameters (\ref{masslessvalues}), we find for this case not only the
Bethe root $\tilde u=0.690849 + 1.5708i$ which corresponds to the
(single) transfer matrix eigenvalue, but also an additional solution
of the Bethe Ansatz equation $\tilde u= 1.4208i$ which does {\bf not}
correspond to this eigenvalue.  As emphasized in the beginning of
Section \ref{sec:generic}, it is difficult to hunt for solutions of
Bethe Ansatz equations, especially for higher values of $N$ and $M$.

Assuming that these two conjectures are correct, it follows that the
Bethe Ansatz yields all $2^{N}$ eigenvalues and no extraneous
solutions for precisely $k=N+1$.  While it is gratifying to obtain all
the eigenvalues, this result is also disappointing, since the
constraint (\ref{constraint}) then implies that the imaginary parts of
the boundary parameters should grow linearly with $N$.

\subsubsection{Ground state}\label{subsubsec:ground}

Although a high value of $k$ is required to obtain all the energy
levels (namely, $k=N+1$), we find that the ground-state energy can be
obtained with a much lower value of $k$.  Indeed, using the parameter
values (\ref{masslessvalues}), we performed a search for the minimum
value of $k$ (for a given value of $N$) where the Bethe
Ansatz reproduces the ground-state energy, up to $N=7$.  Our results
are summarized in Table \ref{table:groundmassless}, which gives in addition to
the value of $k$ also the ground-state energy and the corresponding
Bethe roots.  Our main observation here is that $k=0$ for $N$ odd, and
$k=1$ for $N$ even.  We conjecture that this result is true for generic 
values of the boundary parameters for all
$N$. If correct, then the Bethe Ansatz is practical for investigating 
the ground state in the thermodynamic limit.

We also observe from Table \ref{table:groundmassless} that the shifted
Bethe roots are real for the ground state, as is also the case for the
closed XXZ chain with periodic boundary conditions.  (For higher
values of $k$, the shifted Bethe roots for the ground state are either
real or have imaginary parts $i \pi/2$, as can be seen from Table
\ref{table:massless}.)

Finally, we remark that our numerical results suggest that the Bethe
Ansatz correctly yields $2^{N-1}$ eigenvalues for $k=0$ ($N$ odd), and
$2^{N-1}+ {1\over 2}{N \choose N/2}$ eigenvalues for $k=1$ ($N$ even). 
\footnote{In formulating the latter conjecture, which we have checked
up to $N=8$, a useful reference was \cite{att}.}

\subsection{Massive regime}\label{subsec:massive}

For the case that $\eta$ is purely real, we choose $\theta_{\mp}=0$
and the remaining boundary parameters to be real, thereby making the
transfer matrix manifestly Hermitian. In particular, for the
numerical work presented below, we take the values
\be
\eta &=& 0.3 \,, \quad \alpha_{+} = 0.75 \,, \quad \beta_{+}= -1.2\,, 
\quad \theta_{+} = 0  \,, \non \\
\alpha_{-} &=& 0.25  + (k-1)(0.3)\,, \quad \beta_{-}= 0.5 \,, 
\quad \theta_{-} = 0 \,,
\label{massivevalues}
\ee
which satisfy the constraint (\ref{constraint}) for any
value of $k$.

Our results for the massive regime are very similar to those for the
massless regime.  Indeed, consider Table \ref{table:massive}, which
shows all the $2^{N}$ energy eigenvalues and the corresponding Bethe
roots for values of $N$ ranging from 0 to 4.  As in the massless case,
the minimum value of $k$ for which the Bethe Ansatz reproduces all the
eigenvalues is $k=N+1$.  (We obtained similar results for $N=5$. For 
larger values of $N$, roundoff errors become significant.) 
Moreover, as shown in Table \ref{table:groundmassive}, the Bethe
Ansatz reproduces the ground state energy for $k=0$ for $N$ odd, and
$k=1$ for $N$ even. The corresponding shifted Bethe roots
are purely imaginary.

\section{Special case}\label{sec:special}

We now turn to the special case (\ref{specialcase}), which was first
considered in \cite{Ne1}.  In this case, the boundary terms of the
Hamiltonian (\ref{Hamiltonian}) reduce to
\be
{1\over 2}\sinh \eta \left[ 
\coth \alpha_{-} \tanh \beta_{-} 
\left( \sigma_{1}^{z} - \sigma_{N}^{z} \right)
+ \csch \alpha_{-} \sech \beta_{-}
\left( \sigma_{1}^{x} -\sigma_{N}^{x}\right)
\right] \,.
\ee
We first argue that for this case all the energy eigenvalues are
2-fold degenerate.  Indeed, it is easy to see that the Hamiltonian
commutes with the operator $U$ defined by \footnote{A similar
symmetry operator was invoked in \cite{Al} to argue that an open chain
with {\it diagonal} boundary terms has a two-fold degenerate spectrum
for $N$ odd.  There the argument is simpler, since in that case the
Hamiltonian also commutes with $S^{z}$, while $U$ and $S^{z}$
anticommute.}
\be
U = C\ P ,
\ee
where $C$ is the ``charge conjugation'' operator
\be
C = \prod_{n=1}^{N} \sigma_{n}^{y} \,,
\ee
which satisfies $C^{\dagger} = C$ and $C^{2} = 1$;
and $P$ is the ``parity'' operator \cite{DN}, which satisfies
\be
P\ \sigma_{n}^{j} \ P = \sigma_{N+1-n}^{j} \,,
\ee
as well as $P^{\dagger} = P$ and $P^{2} = 1$.  It follows that also
$U$ is Hermitian and squares to 1.  Hence, $U$ has eigenvalues $\pm
1$.  For $N$ odd, $U$ has an equal number of $+1$ and $-1$
eigenvalues. \footnote{To prove this, it suffices to show that the 
trace of $U$ is zero. For $N$ odd, the parity operator leaves the 
``middle'' spin at site ${1\over 2}(N+1)$ invariant. Hence,
\be
\tr U = \tr_{1 2 \ldots N} U = 
\tr_{{1\over 2}(N+1)} \Big( \sigma_{{1\over 2}(N+1)}^{y} \Big)\ 
\tr' \Big( P \prod_{n \ne {{1\over 2}(N+1)}} \sigma_{n}^{y} \Big) 
= 0 \,, \non
\ee
since the Pauli matrix $\sigma^{y}$ is traceless. (Here $\tr'$ 
denotes trace over all spaces $n \ne {1\over 2}(N+1)$.)} It follows 
that all energy eigenvalues are 2-fold degenerate. In fact, since
$U$ commutes with the full transfer matrix $t(u)$, all the eigenvalues
$\Lambda(u)$ are 2-fold degenerate.

Since for this case there are generally only $2^{N-1}$ distinct
eigenvalues, one expects that all of these eigenvalues can be
reproduced by the Bethe Ansatz with a value of $k < N+1$.  Indeed, as
shown in Table \ref{table:special}, we find significant evidence which
supports the conjecture that the complete set of $2^{N-1}$ eigenvalues
is obtained for $k=0$ (i.e., $M={1\over 2}(N-1)$).  (We obtained
similar results for $N=7$.)

\section{Conclusion}\label{sec:conclude}

Within the range of parameters which we have explored (as detailed 
in Sections \ref{subsec:massless} and \ref{subsec:massive}),
we have found significant numerical evidence for the following
conjectures regarding the Bethe Ansatz solution (\ref{constraint}) -
(\ref{Mvalue}) of the model (\ref{Hamiltonian}):

\begin{itemize}
    
    \item For generic values of the bulk and boundary parameters
    satisfying (\ref{constraint}), the solution yields the complete set of
    $2^{N}$ eigenvalues for $k=N+1$ (i.e., $M=N$).

    \item The solution yields the ground state energy for $k=1$ (i.e.,
    $M={1\over 2}N$) when $N$ is even, and for $k=0$ (i.e., $M={1\over
    2}(N-1)$) when $N$ is odd.  In the massless regime, the shifted Bethe
    roots corresponding to these states are real.

    \item In the special case (\ref{specialcase}) where the spectrum is 
    2-fold degenerate, the Bethe Ansatz solution yields the complete set 
    of $2^{N-1}$ eigenvalues for $k=0$ (i.e., $M={1\over 2}(N-1)$).

\end{itemize}

These results suggest that the Bethe Ansatz solution is both valid and
practical for investigating the ground state (and presumably, also
low-lying excited states) of the model
(\ref{Hamiltonian}) in the thermodynamic limit.  In particular, these
results provide justification for the computations in \cite{Do} of the
thermodynamic limit for the special case (\ref{specialcase}), and
clear the way for analogous computations in the general case. We 
stress, however, that this model has many parameters, other ranges of 
which remain to be explored.

\section*{Acknowledgments}

R.N. is grateful to K. Fabricius and B. McCoy for valuable
discussions, and to the Bologna Section of INFN for its
hospitality.  This work was supported in part by the National Science
Foundation under Grant PHY-0098088 (R.N.), by INFN Grant T012
and by the European Network EUCLID (HPRN-CT-2002-00325) (F.R.).

\bigskip

\noindent
{\bf Note added:}

There is a counterpart of the special case (\ref{specialcase}) for 
even values of $N$, namely,
\be
\alpha_{-}=-\alpha_{+} + \eta \,, \qquad 
\beta_{-}=-\beta_{+} \,, \qquad 
\theta_{+} = \theta_{-} = 0\,, \qquad N =  \mbox{even} \,, \non 
\ee
and hence $k=1$. For this case the spectrum also has degeneracies.
For even values of $N$ up to $N=6$, we find that the Bethe Ansatz solution 
with $M={1\over 2}N$ gives the complete set of 
$2^{N-1}+ {1\over 2}{N \choose N/2}$ distinct eigenvalues.

\begin{table}[htb] 
  \centering
  \begin{tabular}{|c|c|c|c|c|}\hline
    $N$ & $k$ & $M$ & $E$ & shifted Bethe roots $\tilde u_{j}$ \\
    \hline
    0 & 1 & 0 & 0.259617 & -- \\
    1 & 2 & 1 & -0.455662 & 0.278824 \\
      &&& 0.455662 & 0.702627 + $u_{0}$ \\
    2 & 3 & 2 & -1.48624 & 0.124133\,, 0.691313 + $u_{0}$\\
      &&& 0.13202 & 0.334031 $\pm$ 0.186753$i$ \\
      &&& 0.483745 & 0.467053\,, 0.678121 + $u_{0}$ \\
      &&& 0.870471 &  0.55974 +  $u_{0}$\,, 
      0.974697 + $u_{0}$ \\
    3 & 4 & 3 & -2.20345 &  0.0866988\,, 0.373369\,,0.655889 + $u_{0}$\\
      &&& -1.75971 & 0.0794522\,,  0.542856 + $u_{0}$ \,,
      0.939102 + $u_{0}$ \\
      &&& -0.0127446 & 0.240758 $\pm$ 0.139792$i$\,, 
      0.659713 + $u_{0}$ \\
      &&& -0.00644228 & 0.21088, 0.541397 +  $u_{0}$\,, 
      0.935445 + $u_{0}$ \\
      &&& 0.671117 & 0.39724, 0.362185  $\pm$ 0.341578$i$ \\
      &&& 0.878343 & 0.638761 + $u_{0}$\,, 
      0.511044 $\pm$  0.205038$i$ \\
      &&& 1.09747 &0.615979\,,  0.530429 + $u_{0}$\,, 
      0.906019 + $u_{0}$ \\ 
      &&& 1.33542 & 0.468162 + $u_{0}$\,, 
      0.719202 +  $u_{0}$\,, 1.16451 + $u_{0}$ \\
    4 & 5 & 4 & -3.20655 & 0.0656969\,, 0.173673\,, 
    0.505631 +  $u_{0}$\,, 0.869254 + $u_{0}$ \\
      &&& -2.15365 & 0.0641635\,, 0.601829 + $u_{0}$\,, 
      0.390452 $\pm$  0.193885$i$\\
      &&& -1.895 & 0.0609306\,, 0.526013 \,, 
      0.497497 + $u_{0}$\,, 0.847927 +  $u_{0}$  \\
      &&& -1.6139 & 0.0587044 \,, 0.432076 +  $u_{0}$\,, 
      0.672212 +  $u_{0}$\,, 1.10249 + $u_{0}$ \\
      &&& -0.66516 & 0.163048\,, 0.365317 $\pm$  0.193132$i$\,, 
      0.602513 + $u_{0}$ \\
      &&& -0.549331 & 0.145871\,, 0.515342\,, 
      0.497186 + $u_{0}$\,, 0.847232 + $u_{0}$ \\
      &&& -0.345969 & 0.136943\,, 0.431537 + $u_{0}$\,, 
      0.671367 + $u_{0}$ \,, 1.101 + $u_{0}$ \\ 
      &&& 0.259969 & 0.506611 + $u_{0}$\,,  0.87182  + $u_{0}$\,,
      0.160443 $\pm$ 0.150348$i$\\
      &&& 0.842045 & 0.287219, 0.297174  $\pm$  0.310653$i$\,, 
      0.609213 + $u_{0}$ \\
      &&& 0.891361 & 0.497959  + $u_{0}$\,, 
      0.849881 +  $u_{0}$\,, 
      0.370167 $\pm$  0.118987$i$\\
      &&& 0.948984 & 0.285065\,, 0.429407 + $u_{0}$ \,,
      0.668011 + $u_{0}$\,, 1.09495 + $u_{0}$\\
      &&& 1.1934 & 0.429338 $\pm$  0.160247$i$\,, 
      0.373211  $\pm$   0.475013$i$\\
      &&& 1.33834 & 0.56912\,, 0.582163  +  $u_{0}$\,, 
      0.533392  $\pm$ 0.368923$i$\\
      &&& 1.48584 & 0.48142 + $u_{0}$\,, 
      0.807965  +  $u_{0}$\,, 0.649061 $\pm$  0.215203$i$\\
      &&& 1.64707 & 0.733488\,, 0.41686  +  $u_{0}$\,, 
      0.647858  +  $u_{0}$\,, 1.05404  + $u_{0}$ \\
      &&& 1.82255 & 0.367461  +  $u_{0}$\,,
       0.570419 +  $u_{0}$\,, 0.839668  + $u_{0}$\,,
      1.29177 + $u_{0}$  \\ \hline
   \end{tabular}
   \caption{Complete set of $2^{N}$ energy levels and Bethe roots in the
   massless regime, using parameter values (\ref{masslessvalues}).  We
   use the shorthand notation $u_{0} = i \pi/2$.  Without loss of
   generality, we restrict the shifted Bethe roots so that 
   $\Re e\ \tilde u_{j} > 0$ and 
   $-{\pi\over 2} < \Im m\ \tilde u_{j} \le {\pi\over 2}$.}
   \label{table:massless}
\end{table}
\begin{table}[htb] 
  \centering
  \begin{tabular}{|c|c|c|c|c|}\hline
    $N$ & $k$ & $M$ & ground-state energy $E$ & 
    shifted Bethe roots $\tilde u_{j}$\\
    \hline
    1 & 0 & 0 & -2.79413 & -- \\
    2 & 1 & 1 & -1.57715 & 0.0944455 \\
    3 & 0 & 1 & -4.58216 & 0.0973252 \\
    4 & 1 & 2 & -3.3477 & 0.0559452\,, 0.139137 \\
    5 & 0 & 2 & -6.35177 & 0.0572972\,, 0.14122 \\ 
    6 & 1 & 3 & -5.10816 & 0.0402978\,, 0.0893334\,, 0.168789\\ 
    7 & 0 & 3 & -8.11181 & 0.0410605\,, 0.0906602\,, 0.170368\\ \hline
   \end{tabular}
   \caption{Ground-state energy and Bethe roots in the massless regime,
   using parameter values (\ref{masslessvalues}).}
   \label{table:groundmassless}
\end{table}
\begin{table}[htb] 
  \centering
  \begin{tabular}{|c|c|c|c|c|}\hline
    $N$ & $k$ & $M$ & $E$ & shifted Bethe roots $\tilde u_{j}$ \\
    \hline
    0 & 1 & 0 & 0.28216 & -- \\
    1 & 2 & 1 & -0.478182 & 0.274302$i$ \\
      &&& 0.478182 & 1.82287 + $u_{0}$ \\
    2 & 3 & 2 & -1.52836 & 0.123806$i$\,, 1.83483 + $u_{0}$\\
      &&& 0.128052 & 0.184296 $\pm$ 0.327865$i$ \\
      &&& 0.496614 & 0.460762$i$\,, 1.84961 + $u_{0}$ \\
      &&& 0.903692 & 1.44453 + $u_{0}$\,, 2.2506 + $u_{0}$ \\
    3 & 4 & 3 & -2.2733 & 0.0866767$i$\,,  0.369644$i$\,,
      1.87489 + $u_{0}$\\
      &&& -1.80933 &  0.0794024$i$\,, 1.45356 + $u_{0}$\,, 
      2.27091 + $u_{0}$ \\
      &&& -0.00928535 & 0.140292 $\pm$ 0.238892$i$ \,, 1.86994 + $u_{0}$\\
      &&& -0.00463236 & 0.210583$i$\,, 1.45452 + $u_{0}$
      \,, 2.27283 + $u_{0}$ \\
      &&& 0.684253 & 0.387332$i$\,, 0.337297 $\pm$ 0.355183$i$ \\
      &&& 0.903368 & 0.202474 $\pm$ 0.507083$i$\,, 1.89723 + $u_{0}$\\
      &&& 1.12743 & 0.608104$i$\,, 1.46245 + $u_{0}$
      \,, 2.28776 + $u_{0}$\\ 
      &&& 1.38149 & 1.36248 + $u_{0}$\,, 1.78447 + $u_{0}$
      \,, 2.59988 + $u_{0}$ \\
    4 & 5 & 4 & -3.30164 & 0.0657087$i$\,, 0.173726$i$
    \,, 1.47329+ $u_{0}$\,, 2.31403 + $u_{0}$\\
      &&& -2.21969 & 0.0641814$i$\,, 0.192461  $\pm$ 0.388014$i$
      \,,  1.94346 + $u_{0}$\\
      &&& -1.95389 & 0.0609669$i$\,, 0.524409$i$
      \,, 1.47997 + $u_{0}$\,, 2.32536 + $u_{0}$\\
      &&& -1.66252 & 0.0587306$i$\,, 1.36591 + $u_{0}$
      \,, 1.79991 + $u_{0}$\,, 2.62653 + $u_{0}$  \\
      &&& -0.68481 & 0.191746 $\pm$ 0.362798$i$\,, 0.163232$i$
      \,, 1.94239 + $u_{0}$\\
      &&& -0.56579 & 0.146007$i$\,, 0.513987$i$
      \,, 1.4802 + $u_{0}$\,, 2.32576  + $u_{0}$\\
      &&& -0.355209 & 0.137034$i$\,, 1.36599  + $u_{0}$
      \,,  1.80022 + $u_{0}$\,, 2.62701 + $u_{0}$ \\ 
      &&& 0.270674 & 0.150351 $\pm$ 0.160618$i$\,, 
      1.47252+ $u_{0}$\,, 2.31269 + $u_{0}$\\
      &&& 0.87254 & 0.310225 $\pm$ 0.295283$i$
      \,, 0.285824$i$\,, 1.93176 + $u_{0}$\\
      &&& 0.920906 & 0.119052 $\pm$ 0.37073$i$
      \,, 1.47931 + $u_{0}$\,, 2.32448 + $u_{0}$\\
      &&& 0.978875 & 0.285415$i$\,, 1.36633 + $u_{0}$\,,
      1.80148 + $u_{0}$\,, 2.62891 + $u_{0}$\\
      &&& 1.22216 & 0.158173  $\pm$ 0.41471$i$ \,,
      0.46895  $\pm$ 0.364585$i$ \\
      &&& 1.38081 & 0.366909 $\pm$ 0.5330611$i$ 
      \,, 0.565907$i$\,, 1.97405+ $u_{0}$\\
      &&& 1.52904 & 0.212615 $\pm$ 0.654327$i$
      \,, 1.49436 + $u_{0}$\,, 2.34875 + $u_{0}$ \\
      &&& 1.68784 & 0.724014$i$\,, 1.36869 + $u_{0}$
      \,, 1.80952 + $u_{0}$\,, 2.64055 + $u_{0}$\\
      &&& 1.8807 & 1.35068  + $u_{0}$\,, 1.66925 + $u_{0}$\,,
      2.10024 + $u_{0}$\,, 2.91917 + $u_{0}$ \\ \hline
   \end{tabular}
   \caption{Complete set of $2^{N}$ energy levels and Bethe roots in the
   massive regime, using parameter values (\ref{massivevalues}).  We use
   the shorthand notation $u_{0} = i \pi/2$.  Without loss of generality,
   we restrict the shifted Bethe roots so that $\Re e\ \tilde u_{j} > 0$ and
   $-{\pi\over 2} < \Im m\ \tilde u_{j} \le {\pi\over 2}$; or 
   $\Re e\ \tilde u_{j} = 0$ and $0 < \Im m\ \tilde u_{j} \le {\pi\over 2}$.}
  \label{table:massive}
\end{table}
\begin{table}[htb] 
  \centering
  \begin{tabular}{|c|c|c|c|c|}\hline
    $N$ & $k$ & $M$ & ground-state energy $E$  & 
    shifted Bethe roots $\tilde u_{j}$ \\
    \hline
    1 & 0 & 0 & -2.86459 & -- \\
    2 & 1 & 1 & -1.61648 & 0.0941058$i$ \\
    3 & 0 & 1 & -4.71323 & 0.0969713$i$ \\
    4 & 1 & 2 & -3.4432 & 0.0558492$i$\,, 0.138676$i$ \\
    5 & 0 & 2 & -6.53887 & 0.0571973$i$\,, 0.140752$i$ \\ \hline
   \end{tabular}
   \caption{Ground-state energy and Bethe roots in the massive regime,
   using parameter values (\ref{massivevalues}).}
  \label{table:groundmassive}
\end{table}
\begin{table}[htb] 
  \centering
  \begin{tabular}{|c|c|c|c|c|}\hline
    $N$ & $k$ & $M$ & $E$ & shifted Bethe roots $\tilde u_{j}$ \\
    \hline
    1 & 0 & 0 & 0 & -- \\
    3 & 0 & 1 & -2.06361 & 0.0811287 \\
      &&& -0.202938 & 0.228372 \\
      &&& 0.983167 & 1.1779 + $u_{0}$ \\
      &&& 1.28338 & 0.567083$i$ \\
    5 & 0 & 2 & -3.93243 & 0.0518373\,, 0.121304\\
      &&& -2.55868 & 0.0501521\,,  0.259987\\
      &&& -1.61887 & 0.0470911, 1.18204 + $u_{0}$\\
      &&& -1.53627 & 0.114268\,, 0.255101 \\
      &&& -1.3471 & 0.0452599\,, 0.570638$i$ \\
      &&& -0.687493 & 0.10434\,, 1.18027 + $u_{0}$ \\
      &&& -0.474743 & 0.0991988\,, 0.56857$i$ \\
      &&& 0.268219 & 0.119246 $\pm$ 0.149991$i$ \\
      &&& 0.486895 & 0.19328\,, 1.17489 + $u_{0}$ \\
      &&& 0.642915 & 0.17875\,, 0.564459$i$ \\
      &&& 0.954064 & 0.278324 $\pm$ 0.155511$i$ \\
      &&& 1.49874 & 0.407709\,, 1.14897 + $u_{0}$ \\
      &&& 1.65895 & 0.347419\,, 0.557976$i$ \\
      &&& 1.96938 & 0.850516 + $u_{0}$\,, 1.62572 + $u_{0}$ \\
      &&& 2.28162 & 0.551706$i$\,, 1.24328 + $u_{0}$ \\
      &&& 2.3948 & 0.550319$i$\,, 0.950981$i$ \\ \hline
   \end{tabular} 
   \caption{Complete set of $2^{N-1}$ energy levels and Bethe roots for
   the special case (\ref{specialcase}), with $\eta = 0.3 i$,
   $\alpha_{-}=-\alpha_{+}= 0.4 i$, $\beta_{-}=-\beta_{+}= 0.7$,
   $\theta_{+} = \theta_{-} = 0$.  We use the shorthand notation $u_{0} =
   i \pi/2$.  Without loss of generality, we restrict the shifted Bethe
   roots so that $\Re e\ \tilde u_{j} > 0$ and 
   $-{\pi\over 2} < \Im m\ \tilde u_{j} \le {\pi\over 2}$; 
   or $\Re e\ \tilde u_{j} = 0$ and $0 < \Im m\ \tilde u_{j}
   \le {\pi\over 2}$.}
   \label{table:special}
\end{table}

\vfill\eject 

\renewcommand{\theequation}{\arabic{equation}}
\setcounter{equation}{0}
\setcounter{footnote}{0}

\noindent    
{\large ADDENDUM to ``Completeness of the Bethe Ansatz solution of the 
open XXZ chain with nondiagonal boundary terms''}

\bigskip

In [1] (to which we refer hereafter by I), we find significant
numerical evidence that for the open XXZ quantum spin chain Hamiltonian 
(I1.1) with bulk and boundary parameters satisfying the constraint (I1.2)
\be
\alpha_{-} + \beta_{-} + \alpha_{+} + \beta_{+} = \pm (\theta_{-} - 
\theta_{+}) + \eta k \,,
\label{addconstraint}
\ee
the Bethe Ansatz solution (I1.3) - (I1.8), (I2.9) gives the 
complete set of $2^{N}$ eigenvalues for $k=N+1$; and this solution gives 
only some (but not all) of the eigenvalues for $1-N \le k < N+1$.
Here we conjecture that a simple generalization of this
Bethe Ansatz solution gives the {\it complete} set of eigenvalues 
for {\it all} values of $k$ in the interval $-(N+1) \le  k \le N+1$. 

Indeed, consider the following two expressions, distinguished by 
$\pm$, for the eigenvalues of the transfer matrix (I2.4):
\be
\Lambda_{\pm}(u) = h_{\pm}(u) {Q_{\pm}(u-\eta)\over Q_{\pm}(u)} 
+ h_{\pm}(-u-\eta) {Q_{\pm}(u+\eta)\over Q_{\pm}(u)}  \,,
\label{addopeneigenvalues} 
\ee
where $h_{\pm}(u)$ are given by
\be 
h_{\pm}(u) &=& -\sinh^{2N}(u+\eta){\sinh(2u+2\eta)\over \sinh(2u+\eta)} 
\non \\
&\times& 4 \sinh(u \pm \alpha_{-}) \cosh(u \pm \beta_{-}) 
\sinh(u \pm \alpha_{+}) \cosh(u \pm \beta_{+}) 
\,; \label{addhfinal} 
\ee
and $Q_{\pm}(u)$ are given by
\be
Q_{\pm}(u) = \prod_{j=1}^{M_{\pm}} \sinh(u - u_{j}^{\pm}) 
\sinh(u + u_{j}^{\pm} + \eta) \,,
\label{addopenQ}
\ee 
where 
\be
M_{\pm}={1\over 2}(N-1 \pm k) \,.
\label{addMvalue}
\ee
The corresponding Bethe Ansatz equations are 
\be
{h_{\pm}(u_{j}^{\pm})\over h_{\pm}(-u_{j}^{\pm}-\eta)} = 
-{Q_{\pm}(u_{j}^{\pm}+\eta)\over Q_{\pm}(u_{j}^{\pm}-\eta)} \,, 
\qquad j = 1 \,, \ldots \,, M_{\pm} \,.
\label{openBAeqs}
\ee
The Bethe Ansatz solution considered earlier [2,1] corresponds
to the above expressions with the plus ($+$) sign.

We conjecture that for a given set of bulk and boundary parameters 
satisfying (\ref{addconstraint}) with $|k| \le N+1$ and $k$ odd (even)
for $N$ even (odd), the eigenvalues $\Lambda_{+}(u)$ and
$\Lambda_{-}(u)$ together give the complete set of 
eigenvalues of the transfer matrix. That is, for $k=N+1$,
all the eigenvalues are given by $\Lambda_{+}(u)$ (as found in I);
for $k=-(N+1)$, all the eigenvalues are given by $\Lambda_{-}(u)$;
and for $-(N+1) < k < N+1$, both $\Lambda_{+}(u)$ and 
$\Lambda_{-}(u)$ are needed to obtain the complete set of 
eigenvalues.

This conjecture is supported by significant numerical evidence,
obtained (as in I) using `McCoy's method'. For instance, for
$N=6$ and $k=1$, we find that 42 eigenvalues are given by $\Lambda_{+}(u)$
and 22 eigenvalues are given by $\Lambda_{-}(u)$; together, they
give the complete set of $2^{6} = 64$ eigenvalues.

\end{document}